\documentclass[twocolumn,showpacs,preprintnumbers,floatfix,
               amsmath,amssymb]{revtex4}
\usepackage{longtable}
\usepackage{graphics}
\usepackage{epsfig}

\begin{document}

\title{The Shear Viscosity and Thermal Conductivity of Nuclear Pasta}

\author{C. J. Horowitz}\email{horowit@indiana.edu} 
\affiliation{Nuclear Theory Center and Department of Physics, 
             Indiana University,  Bloomington, IN 47405}
\author{D. K. Berry}\email{dkberry@indiana.edu}
\affiliation{University Information Technology Services,
             Indiana University, Bloomington, IN 47408}


\date{\today}

\begin{abstract}
We calculate the shear viscosity $\eta$ and thermal conductivity $\kappa$ of a nuclear pasta phase in neutron star crusts.  This involves complex non-spherical shapes.  We use semiclassical molecular dynamics simulations involving 40,000 to 100,000 nucleons.  The viscosity $\eta$ can be simply expressed in terms of the height $Z^*$ and width $\Delta q$ of the peak in the static structure factor $S_p(q)$.  We find that $\eta$ increases somewhat, compared to a lower density phase involving spherical nuclei, because $Z^*$ decreases from form factor and ion screening effects.  However, we do not find a dramatic increase in $\eta$ from non-spherical shapes, as may occur in conventional complex fluids.
\end{abstract}
\maketitle
%
\section{Introduction}
\label{intro}
The viscosity of dense matter can dampen the oscillations of neutron stars.  High multipolarity p-mode oscillations may impact the pulse shapes of some radio pulsars \cite{pulsars}.  For p-modes, the primary restoring force is the pressure, and the modes may be damped by the shear viscosity of the neutron star crust \cite{chugunov}.  Inertial or r-modes of a rotating neutron star may radiate gravitational waves and could limit the spin period \cite{rmodes}.  For these modes the coriolis force provides the primary restoring force and they may be damped by the shear, and or bulk, viscosities of dense matter through out the star. 

Flowers and Itoh \cite{flowers} and Nandkumar and Pethick \cite{nandkumar} have calculated the shear viscosity of neutron star crust matter in the liquid phase.  While recently, Chugunov and Yakovlev have calculated the shear viscosity of both the solid and liquid phases \cite{chugunov}.  They assume the viscosity is dominated by the momentum carried by degenerate electrons and that this is limited by electron-ion scattering.   At high densities in the neutron star core, the shear viscosity comes from momentum carried by nucleons and this is limited by NN scattering.  See for example Ref. \cite{benhar}.

The thermal conductivity of the neutron star crust helps determine the temperature profile of the star and is very important for the cooling time of the crust.  Recently, crust cooling times have been observed for two neutron stars after extended outbursts \cite{cooling}.  We have calculated the thermal conductivity of the outer crust including the possible role of impurities \cite{thermalcon}.  The thermal conductivity could also be important for setting the ignition conditions for carbon superbursts \cite{superbursts}.  

At intermediate densities, around 10$^{14}$ g/cm$^3$ just below nuclear density, matter may form complex nuclear pasta phases \cite{pasta}.  Competition between short range attractive nuclear, and long range repulsive Coulomb, interactions can lead to clusters with many different non-spherical shapes including long rods or flat plates.  Because pasta may form at high densities, it could represent as much as half the mass of the neutron star crust.  The complex shapes in nuclear pasta have sizes of tens of Fermis.  This is comparable to the wavelength of neutrinos in Supernova explosions.  Therefore, coherent scattering from the nuclear pasta shapes may impact the opacity of Supernova neutrinos \cite{pastaopacity}.

In the laboratory, complex fluids with large non-spherical molecules can have shear viscosities that are much larger than for simple fluids.  How might the complex pasta shapes influence the shear viscosity of nuclear pasta?  Is there a drastic increase in viscosity that is similar to that seen for laboratory complex fluids?  We are not aware of any previous calculations of the shear viscosity of nuclear pasta.  Although Chugunov and Yakovlev have calculated the shear viscosity for similar densities \cite{chugunov}, they apparently assumed spherical nuclei.           

In this paper, we calculate the shear viscosity and thermal conductivity using a semiclassical model for the nuclear pasta \cite{pasta1} and molecular dynamics (MD) simulations with from 40,000 to 100,000 nucleons \cite{pasta2}.  Our goal is to gain some simple insight into how transport properties, such as the shear viscosity or thermal conductivity, depend on the sizes and shapes of the clusters.  Furthermore, we wish to know how transport properties might change as one goes from isolated spherical nuclei at low densities, to complex pasta phases, and then to uniform nuclear matter at high densities.  One important advantage of our semiclassical model is that it can be directly applied in all of these density regimes.  

We wish to emphasize the importance of simple qualitative insight.   Astrophysicists request very detailed transport properties, such as the shear viscosity, even though there are large remaining uncertainties in much more basic dense matter properties.  It may be unrealistic to predict the exact sizes and shapes of the pasta clusters.  Instead, we hope to gain qualitative insight into how pasta shapes may impact the shear viscosity or other properties.  

The manuscript is organized as follows. In Sec.~\ref{formalism} we review the simple semiclassical model, explain the calculation of the shear viscosity, and provide some details of the molecular dynamics simulations.  In Sec.~\ref{results} we present results for the static structure factors and use these to calculate the shear viscosity and thermal conductivity.  Finally, conclusions and future directions are presented in Sec.~\ref{conclusions}.

\section{Formalism}
\label{formalism}
In Subsection \ref{pasta} we review our semiclassical model that can describe nuclear pasta phases in a flexible way.  Next, in Subsection \ref{viscosity} we describe the calculation of the shear viscosity based on molecular dynamics simulations.

\subsection{Semiclassical Nuclear Pasta Model}
\label{pasta}

We briefly describe our semiclassical model that while simple, contains the essential physics of competing interactions consisting of a short-range nuclear attraction and a long-range Coulomb repulsion.  This competition can generate complex pasta shapes.  The impossibility to simultaneously minimize all elementary interactions is known in condensed-matter circles as {\it
frustration}. The complex physics of frustration, along with many
other details of the model, may be found in Refs.~\cite{pasta1,pasta2}.  Here only a brief review of the most essential features of the model is presented. We model a charge-neutral system of electrons, protons, and neutrons. The electrons are assumed to be noninteracting and thus are described as a degenerate free Fermi gas at a number density identical to that of the protons ({\it i.e.,} $n_e\!=\!\rho_p$).  The nucleons, on the other hand, interact classically via a
nuclear-plus-Coulomb potential.  However, the use of an effective
temperature and effective interactions are used to simulate effects
associated with quantum zero-point motion.    While simple, the model displays the essential physics of frustration, namely, nucleons clustering into pasta but the size of the clusters limited by the Coulomb repulsion, in a transparent form. Moreover, one may study the evolution of the system through the low density, pasta, and high density phases within a single microscopic model. Finally, the model facilitates simulations with a large numbers of particles, a feature that is essential to estimate and control finite-size effects and to reliably study the long wavelength response of the system.

The total potential energy $V_{\rm tot}$ of the system consists of a
sum of two-body interactions
\begin{equation}
 V_{tot}=\sum_{i<j} V(i,j) \;,
\label{vtot}
\end{equation}
where the ``elementary'' two-body interaction is given as follows:
\begin{equation}
V(i,j) = a e^{-r_{ij}^{2}/\Lambda} + \Big[b+c\tau_z(i)\tau_z(j)\Big]
       e^{-r_{ij}^{2}/2\Lambda}+V_{\rm c}(i,j)\;.
\label{v}
\end{equation}
Here the distance between the particles is denoted by $r_{ij}=|{\bf
r}_i\!-\!{\bf r}_j|$ and $\tau_z$ represents the nucleon isospin
projection ($\tau_z\!=\!+\!1$ for protons and $\tau_z\!=\!-\!1$ for
neutrons).  The two-body interaction contains the characteristic
intermediate-range attraction and short-range repulsion of the
nucleon-nucleon force. Further, an isospin dependence has been
incorporated in the potential to ensure that while pure neutron matter
is unbound, symmetric nuclear matter is appropriately bound. Indeed,
the four model parameters ($a$, $b$, $c$, and $\Lambda$) introduced in
Eq.~({\ref v}) have been adjusted in Ref.~\cite{pasta1} to reproduce
the following bulk properties: a) the saturation density and binding
energy per nucleon of symmetric nuclear matter, b) (a reasonable value
for) the binding energy per nucleon of neutron matter at saturation
density, and c) (approximate values for the) binding energy of a few
selected finite nuclei. All these properties were computed via a
classical Monte Carlo simulation with the temperature arbitrarily
fixed at 1 MeV. The parameter set employed in all previous and present
calculations is displayed in Table~\ref{Table1}.
Finally---and critical for pasta formation---a screened Coulomb
interaction of the following form is included:
\begin{equation}  
V_{\rm c}(i,j)=\frac{e^{2}}{r_{ij}}e^{-r_{ij}/\lambda}
                 \tau_p(i)\tau_p(j) \;,
\label{vc}
\end{equation}
where $\tau_p\!\equiv\!(1\!+\!\tau_z)/2$ and $\lambda$ is the
screening length that results from the slight polarization of the
electron gas. The relativistic Thomas-Fermi screening length is given
by
\begin{equation}
 \lambda=\frac{\pi^{1/2}}{2e} 
 \left(k_{\rm F}\sqrt{k_{\rm F}^2+m_e^2}\right)^{-1/2}
 \hspace{-0.2cm}\;,
 \label{lambda}
\end{equation}
where $m_e$ is the electron mass, the electron Fermi momentum has
been defined by $k_{\rm F}\!=\!(3\pi^2n_e)^{1/3}$, and $n_e$ is the electron density~\cite{pasta1}.  Unfortunately, while the screening length $\lambda$, defined above, is smaller than the
length $L$ of our simulation box, it is not significantly
smaller.  Hence, following a prescription introduced in
Ref.~\cite{pasta1} in an effort to control finite-size effects, the
value of the screening length is arbitrarily decreased to
$\lambda\!=\!10$~fm.  This decrease in the screening length could slightly increase the size of clusters because it somewhat reduces Coulomb repulsion.  However, we do not expect this to be a large change because $\lambda$ is still larger than, or comparable to, the size of clusters.  In the future, MD simulations should be performed using the full physical screening length.

The simulations are carried out with both a fixed number of particles
$A$ and a fixed density $\rho$. The simulation volume is then simply
given by $V\!=\!A/\rho$. To minimize finite-size effects periodic
boundary conditions are used.  To carry out molecular dynamics simulations the
trajectories of all of the particles in the system are determined by
simply integrating Newton's laws of motion, albeit for a large
number of particles (up to 100,000 in the present case) using the
velocity-Verlet algorithm~\cite{verlet}. To start the
simulations, initial positions and velocities must be specified for
all the particles in the system. The initial positions are randomly
and uniformly distributed throughout the simulation volume while the
initial velocities are distributed according to a Boltzmann
distribution at temperature $T$.  As the velocity-Verlet is an
energy---not temperature---conserving algorithm, kinetic and potential
energy continuously transform into each other.  To prevent these
temperature fluctuations, the velocities of all the particles are
periodically rescaled to ensure that the average kinetic energy per 
particle remains fixed $(3/2)k_{\rm B}T$.

In summary, a classical system has been constructed with a total
potential energy given as a sum of two-body, momentum-independent
interactions as indicated in Eq.~(\ref{v}). Expectation values of any
observable of interest may be calculated as a suitable time average
using particle trajectories generated from molecular dynamics
simulations.

We comment on the classical nature of our simulations.  Correlations from Fermi statistics are not explicitly included.  However, some effects of Pauli exclusion are implicitly included by incorporating short range repulsion in Eq. (\ref{vc}) and adjusting the parameters in Table \ref{Table1} to reproduce the saturation density and binding energy of nuclear matter.  In this paper, we focus on the electron-proton response at long wavelengths.  This is dominated by correlations between clusters.  These clusters are heavy, since they involve many nucleons, and their thermal de Broglie wavelengths are much shorter than the inter-cluster spacing.  Therefore, we expect our classical approximation to accurately reproduce the long wavelength response. 

\subsection{Shear Viscosity}
\label{viscosity}
Following Chugunov and Yakovlev \cite{chugunov}, we assume the shear viscosity $\eta$ is dominated by momentum carried by electrons and that this is limited by electron-ion scattering,
\begin{equation}
\eta = \frac{\pi v_F^2 n_e}{20Z_{ion}\alpha^2 \Lambda_{ei}}\, .
\end{equation}
Chugunov et al work in ion coordinates with $Z_{ion}$ the ion charge, the electron Fermi velocity is $v_F\approx 1$, and $n_e$ is the electron density.  The Coulomb logarithm $\Lambda_{ei}$ describes electron ion scattering \cite{chugunov},
\begin{equation}
\Lambda_{ei}=\int_0^{2k_F} \frac{dq}{q} \frac {F(q)^2}{\epsilon(q)^2} S_{ion}(q) (1-\frac{q^2}{4k_F^2})(1-\frac{v_F^2q^2}{4k_F^2})\, .
\end{equation}
Here $\epsilon(q)$ is the static longitudinal dielectric function of the electron gas \cite{dielectric} ($\epsilon(q)\approx 1$ except at low momentum transfer $q$).  The static structure factor $S_{ion}(q)$ describes correlations between ions and can be calculated from the ion density-density correlation function, see below.  The nuclear form factor $F(q)$ is,
\begin{equation}
F(q)=\frac{1}{Z_{ion}} \int d^3r {\rm e}^{i{\bf q}\cdot{\bf r}} \rho_p(r)\, ,
\label{ff}
\end{equation}
where $\rho_p(r)$ is the proton density inside one ion ($F(q=0)=1$).

As the density increases, and the nuclei start to strongly interact to form complex pasta shapes, one may no longer be able to effectively use ion coordinates.  Therefore we work directly in the nucleon coordinates instead of using the ion coordinates.  At low densities the nucleons form clusters in our model that are equivalent to nuclei.  As a result our calculation in the nucleon coordinates, at low densities, will reproduce a calculation in the ion coordinates.  We replace the Coulomb logarithm describing electron-ion scattering with one describing electron-proton scattering $\Lambda_{ep}$,
\begin{equation}
Z_{ion} \Lambda_{ei} \rightarrow \Lambda_{ep}\, ,
\end{equation}  
and write,
\begin{equation}
\eta =\frac{\pi v_F^2 n_e}{20 \alpha^2 \Lambda_{ep}}\, .
\label{eta}
\end{equation}
The electron-proton Coulomb logarithm is,
\begin{equation}
\Lambda_{ep} = \int_0^{2k_F} \frac{dq}{q}\frac{f_{sn}(q)^2}{\epsilon(q)^2} S_p(q) (1-\frac{q^2}{4k_F^2})(1-\frac{v_F^2q^2}{4k_F^2})\, .
\label{coullog}
\end{equation}
Here $f_{sn}(q)$ is the single nucleon form factor (charge distribution of the proton) that we approximate $f_{sn}(q)=1$.  The static structure factor $S_p(q)$ describes correlations between protons and is calculated from the density-density correlation function
\begin{equation}
S_p(q)=\langle \rho_p(q)^*\rho_p(q) \rangle\, ,
\label{Sp(q)}
\end{equation}
Here $\rho_p(q)$ is the proton density
\begin{equation}
\rho_p(q)=\frac{1}{\sqrt{N_p}} \sum_{i=1}^{N_p} {\rm e}^{i{\bf q}\cdot{\bf r}_i(t)}\, ,
\label{spqeq}
\end{equation}
for a simulation with $N_p$ protons at positions ${\bf r}_i(t)$ and the statistical average in Eq. \ref{Sp(q)} is calculated as an average over the time $t$ in the MD simulation.

Our model automatically includes the correlations between nucleons that go into forming ions of charge $Z_{ion}$.  Furthermore, the model also includes the distributions of protons inside a nucleus that is described by the form factor $F(q)$.  In the limit where $Z_{ion}$ protons correlate into each cluster and there are only weak correlations between the clusters one would have 
\begin{equation}
S_p(q)\approx Z_{ion} F(q)^2\, ,
\end{equation}
and
\begin{equation}
S_{ion}(q)\approx 1
\end{equation}
so that
\begin{equation}
Z_{ion} S_{ion}(q) F(q)^2 \approx S_p(q)\, .
\end{equation}

In this paper we are interested in transport properties of electrons.  These depend on the proton static structure factor $S_p(q)$.  However for comparison, we also consider the static structure factor for neutrons $S_n(q)$.  This is important for neutrino transport.  Neutrinos in a supernova scatter from the weak charge density.  In the Standard model the weak charge of a neutron is much larger than that of a proton.  Therefore neutrino scattering can be described with the static structure factor for neutrons $S_n(q)$, see for example \cite{pasta1}\cite{pastaopacity}.
\begin{equation}
S_n(q)=\langle \rho_n(q)^*\rho_n(q) \rangle\, ,
\label{Sn(q)}
\end{equation}
Here $\rho_n(q)$ is the neutron density
\begin{equation}
\rho_n(q)=\frac{1}{\sqrt{N_n}} \sum_{i=1}^{N_n} {\rm e}^{i{\bf q}\cdot{\bf r}_i(t)}\, ,
\end{equation}
where the sum runs over $N_n$ neutrons at positions ${\bf r}_i(t)$.  We will calculate $S_p(q)$ and $S_n(q)$ in Section \ref{results}.

\begin{table}
\caption{Model parameters used in the calculations.}
 \begin{ruledtabular}
 \begin{tabular}{cccc}
   $a$     & $b$     & $c$    & $\Lambda$ \\
   \hline
   110 MeV & -26 MeV & 24 MeV & 1.25 fm$^2$
 \label{Table1}
 \end{tabular}
\end{ruledtabular}
\end{table}

\begin{table}
\caption{Simulation parameters and shear viscosity results:  the baryon density is $\rho$, the total number of particles in the simulation $A$, and $T_{tot}$ is the total simulation time.  The height of the peak in $S_p(q)$ is $Z^*$, at location $q^*$ and full width $\Delta q$.  The Coulomb logarithm, Eq. \ref{coullog}, is $\Lambda_{ep}$ and $Z^*\Delta q/q^*$ provides a simple approximation to $\Lambda_{ep}$, see text.  The shear viscosity (in units of $10^{10}$ Pascal-seconds) is $\eta$. }
 \begin{ruledtabular}
 \begin{tabular}{ccccccccc}
   $\rho$ & $A$ & $T_{tot}$ & $Z^*$ & $q^*$ & $\Delta q$ & $\frac{Z^* \Delta q}{q^*}$ & $\Lambda_{ep}$ & $\eta$  \\
   fm$^{-3}$ & & fm/c & & fm$^{-1}$ & fm$^{-1}$ & & & 10$^{10}$ Pa-s\\
   \hline
0.01 & 40000 & $1.3\times 10^6$ & 34.4 & 0.28 & 0.108 & 13.3 & 13.7 & $5.3$\\
0.025 & 100000& 52000 & 38.6 & 0.31 & 0.077 & 9.5 &10.4 & $17$\\
0.05 & 100000 & 28000 & 20.3 & 0.38 & 0.090 & 4.9 & 5.8 & $63$ \\
 \label{Table2}
 \end{tabular}
\end{ruledtabular}
\end{table}

\section{Results}
\label{results}
In this section we present MD simulation results for static structure factors, shear viscosity, and thermal conductivity.

\subsection{MD Simulations}
\label{MDsim}

\begin{figure}[ht]
\begin{center}
\includegraphics[width=2.75in,angle=0,clip=false]{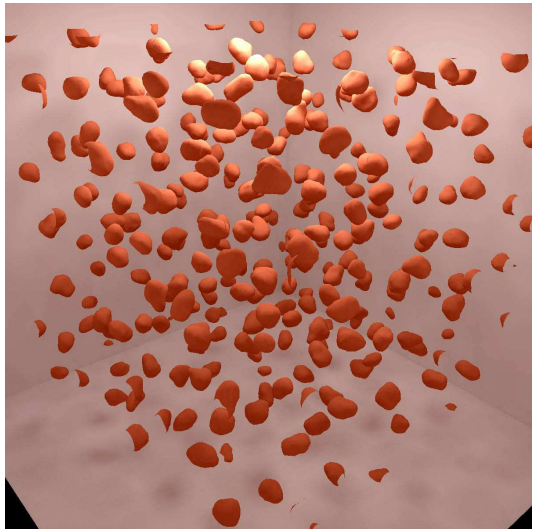}
\caption{(Color online) The 0.03 fm$^{-3}$ proton density isosurface for one configuration of 40,000 nucleons at a density of 0.01 fm$^{-3}$.  The simulation volume is a cube 159 fm on a side.}
\label{Fig1}
\end{center}
\end{figure}

\begin{figure}[ht]
\begin{center}
\includegraphics[width=2.75in,angle=0,clip=false]{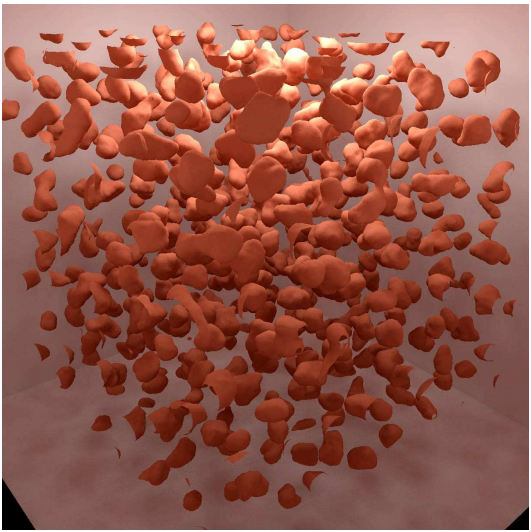}
\caption{(Color online) The 0.03 fm$^{-3}$ proton density isosurface for one configuration of 100,000 nucleons at a density of 0.025 fm$^{-3}$.  The simulation volume is a cube 159 fm on a side.}
\label{Fig2}
\end{center}
\end{figure}

The MD simulations of ref. \cite{pasta1} were for a temperature of 1 MeV, a proton fraction of $Y_p=0.2$, and baryon densities $\rho$ of 0.01, 0.025, and 0.05 fm$^{-3}$, See Table \ref{Table2}.  Note that the proton fraction $Y_p=0.2$ is intermediate between the higher proton fraction expected in a supernova core during the in fall phase and the lower proton fraction expected for neutron star crust in beta equilibrium.  Figure \ref{Fig1} shows the 0.03 fm$^{-3}$ iso-surface of the proton density for one configuration of the $\rho=0.01$ fm$^{-3}$ simulation.  At this density, all of the protons and most of the neutrons are clustered into nuclei.  There is also a low density neutron gas between the clusters which is not shown.   Figure \ref{Fig2} shows the proton density at the higher baryon density $\rho=0.025$ fm$^{3}$.  Now the clusters are larger (larger mass nuclei) and are closer together.  The nuclei are larger because the higher density electron gas cancels more of the Coulomb repulsion.  This allows nuclei to form with more protons.

\begin{figure}[ht]
\begin{center}
\includegraphics[width=2.75in,angle=0,clip=false]{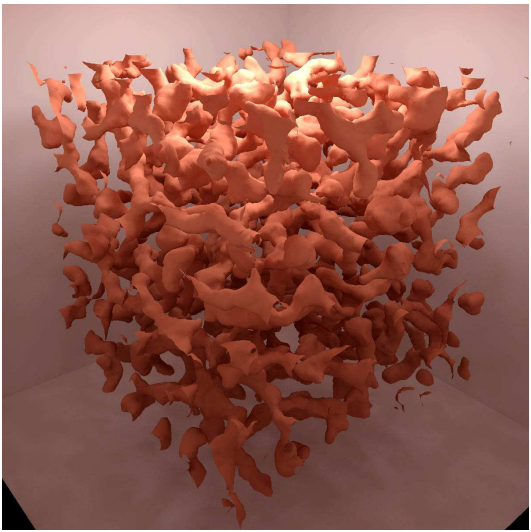}
\caption{(Color online) The 0.03 fm$^{-3}$ proton density isosurface for one configuration of 100,000 nucleons at a density of 0.05 fm$^{-3}$.  The simulation volume is a cube 126 fm on a side.}
\label{Fig3}
\end{center}
\end{figure}

Figure~\ref{Fig3} shows the proton density for one configuration of 100,000 nucleons at a density of $0.05$ fm$^{-3}$.  The clusters are now seen to have very elongated shapes.  The low density neutron gas between these clusters is not shown.  These elongated spaghetti like shapes are one example of a nuclear pasta phase.  Note that the shapes are not straight rods but instead they bend and branch inside the simulation volume.  Because of the periodic boundary conditions, the shapes extend out one side and back in another side of the simulation volume.

\subsection{Static Structure Factor Results}
\label{S(q)results}

\begin{figure}[ht]
\begin{center}
\includegraphics[width=2.75in,angle=270,clip=false]{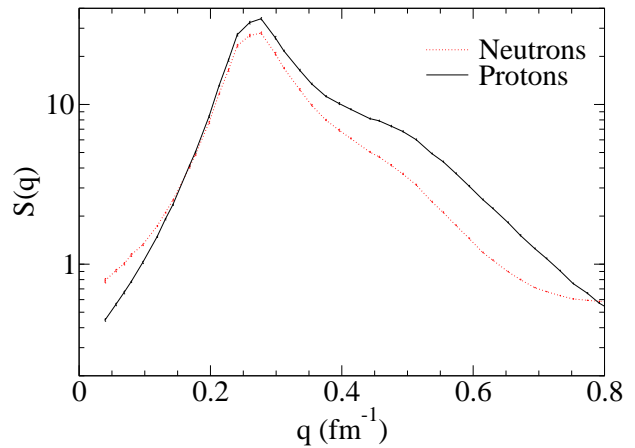}
\caption{(Color online) The static structure factor $S_p(q)$ for protons (solid black line), and for neutrons $S_n(q)$ (dotted red line) versus momentum transfer $q$ at a baryon density of 0.01 fm$^{-3}$.}
\label{Fig4}
\end{center}
\end{figure}

We calculate the static structure factors $S_p(q)$ for protons and $S_n(q)$ for neutrons at $\rho=0.01$ fm$^{-3}$ by averaging over 16000 configurations of the 40000 nucleons where each configuration is separated from the next by a simulation time of 20 fm/c.  Note that calculating the static structure factor by averaging over a finite time $16000 \times 20 =320000$ fm/c may slightly under estimate the height of the peaks in $S_p(q)$ and $S_n(q)$ compared to averages over much longer times.  We consider momentum transfers ${\bf q}$,
\begin{equation}
{\bf q}=\frac{2\pi}{L}(n_x, n_y, n_z)\, ,
\end{equation}
with $L$ the length of the simulation volume and $n_x$, $n_y$, $n_z$ are integers.  We average over the directions of ${\bf q}$.  Figure \ref{Fig4} shows $S_p(q)$ and $S_n(q)$ at $\rho=0.01$ fm$^{-3}$.

Our results for $S_n(q)$ agree well with the original results for $S(q)$ presented in ref. \cite{pasta2}.  Note that in ref. \cite{pasta2}, $S(q)$ was calculated from the Fourier transform of the radial distribution function $g(r)$ (because this was available).  However this involved some uncertainty from the assumed form of $g(r)$ at large distances $r$.  Our results for $S_p(q)$ are new.  This was not calculated in ref. \cite{pasta2}.

The static structure factor $S_n(q)$ is larger than $S_p(q)$ at small $q$.  This is because long wavelength fluctuations in the neutron density need not feel an electrostatic restoring force.  Therefore they are not strongly screened.  In contrast all long wavelength fluctuations in the proton density are strongly screened and this makes $S_p(q)$ small at small $q$.  This has been discussed in ref. \cite{hetero}.  See also ref. \cite{sawyers}.

\begin{figure}[ht]
\begin{center}
\includegraphics[width=2.75in,angle=270,clip=false]{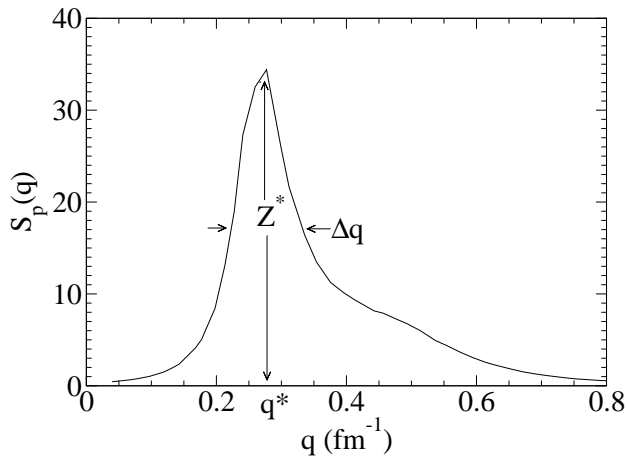}
\caption{The static structure factor $S_p(q)$ for protons versus momentum transfer $q$ at a baryon density of 0.01 fm$^{-3}$ (note the linear scale).  The height of the peak in $S_p(q)$ is $Z^*$, the full width at half maximum is $\Delta q$, and the peak position is $q^*$.}
\label{Fig4a}
\end{center}
\end{figure}

At intermediate $q$, there is a large peak in both $S_p(q)$ and $S_n(q)$.  This corresponds to elastic electron-nucleus (for $S_p(q)$) or neutrino-nucleus (for $S_n(q)$) scattering.  The cross section for coherent elastic electron-nucleus scattering goes like the atomic number squared $Z^2$. However, the structure factor $S_p(q)$ is normalized per proton (instead of per ion).  Therefore, the cross section per proton is proportional to $Z$.  One can think of the height of the peak in $S_p(q)$ as being the effective number of protons $Z^*$ that the electron scatters from.  This effective number of protons $Z^*$ is reduced from the actual number of protons in a nucleus because of the form factor $F(q)$, Eq. \ref{ff}.  The form factor leads to a large reduction at high $q$.  In addition $Z^*$ is reduced by the screening effects of other ions.  This reduction is very large at small $q$.  Finally correlations between ions can increase $Z^*$ somewhat, see ref. \cite{pasta2}.  This corresponds to $S_{ion}(q)>1$.  We characterize the peak in $S_p(q)$ by its height $Z^*$, its full width at half maximum $\Delta q$, and its location $q^*$.  These values are illustrated in Fig. \ref{Fig4a} and collected in Table \ref{Table2}.  

The system is neutron rich.  Therefore one might expect the peak in $S_n(q)$, representing an effective number of neutrons $N^*$, to be larger than the peak in $S_p(q)$.  However this is not the case.  A neutrino will scatter from neutron density contrasts.  The more or less uniform neutron gas, that is also present outside the nuclei, leads to a reduction in contrast.  One only scatters from the difference in inside to outside neutron density.  As a result the peak in $S_n(q)$ is smaller than the peak in $S_p(q)$ even though the nuclei are very neutron rich.        

\begin{figure}[ht]
\begin{center}
\includegraphics[width=2.75in,angle=270,clip=false]{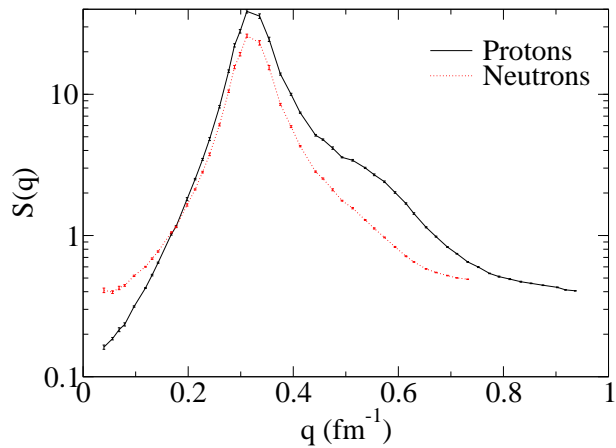}
\caption{(Color online) The static structure factor $S_p(q)$ for protons (solid black line), and for neutrons $S_n(q)$ (dotted red line) versus momentum transfer $q$ at a density of 0.025 fm$^{-3}$.}
\label{Fig5}
\end{center}
\end{figure}

We calculate the static structure factors $S_p(q)$ and $S_n(q)$ at $\rho=0.025$ fm$^{-3}$ by averaging over 10500 configurations of the 100000 nucleons, of the simulation shown in Fig. \ref{Fig2}.   Each configuration is separated from the next by 20 fm/c.  Figure \ref{Fig5} shows $S_p(q)$ and $S_n(q)$ at $\rho=0.025$ fm$^{-3}$.   Again there is a large peak in $S_p(q)$ that is slightly higher than the peak at $\rho=0.01$ fm$^{-3}$, in Fig. \ref{Fig4}.  The nuclei at $\rho=0.025$ fm$^{-3}$ have about twice the number of nucleons $A\approx 200$ as those at $\rho=0.01$ fm$^{-3}$ which have $A\approx 100$, see ref. \cite{pasta2}.  Nevertheless $Z^*$ is only slightly higher at $\rho=0.025$ fm$^{-3}$ than at $\rho=0.01$ fm$^{-3}$.  This is because screening from the other ions is more effective and extends to higher $q$ at this higher density.  This limits the peak in $S_p(q)$ from the left (for low $q$).  Also the nuclear form factor $F(q)$ falls more quickly with $q$ because the nuclei are now larger.  This limits the peak in $S_p(q)$ from the right (for high $q$).  As a result the peak is narrower (has a smaller $\Delta q$) at a density of 0.025 than at 0.01 fm$^{-3}$.

\begin{figure}[ht]
\begin{center}
\includegraphics[width=2.75in,angle=270,clip=false]{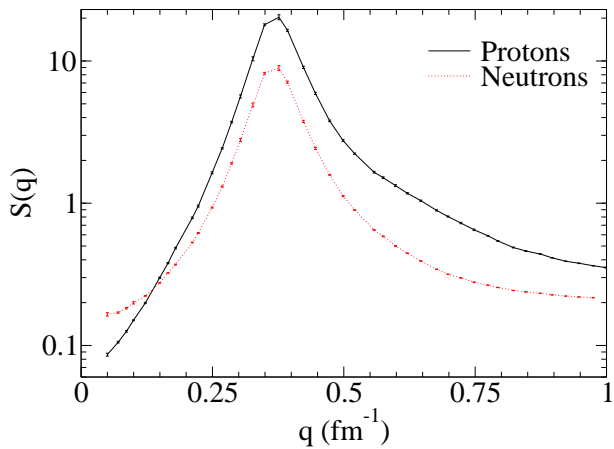}
\caption{(Color online) The static structure factor $S_p(q)$ for protons (solid black line), and for neutrons $S_n(q)$ (dotted red line) versus momentum transfer $q$ at a density of 0.05 fm$^{-3}$.  At this density the system is in a nuclear pasta phase.}
\label{Fig6}
\end{center}
\end{figure}

Finally we calculate $S_p(q)$ and $S_n(q)$ at a density 0.05 fm$^{-3}$ by averaging over 19400 configurations of 100000 nucleons.   Again each configuration is separated by 20 fm/c.  At this density the system is in a nuclear pasta configuration with complex rod-like shapes, see Fig. \ref{Fig3}.   Figure \ref{Fig6} shows $S_p(q)$ and $S_n(q)$ at 0.05 fm$^{-3}$.  There is still a peak in $S_p(q)$.  However it is not as high as at lower densities.  Each piece of pasta contains a very large number of nucleons.  However, they are packed closely together and this allows screening from other pieces of pasta to be very effective.  This greatly reduces $S_p(q)$.  Figure \ref{Fig6} shows no qualitative effect from the non-sperical pasta shapes.  Instead $S_p(q)$ has very similar behavior for round nuclei in Figs. \ref{Fig4},\ref{Fig5} or long rods in Fig. \ref{Fig6}.  Note that we have averaged over the directions of ${\bf q}$.  The system can have large pasta shapes with very many nucleons.  However, it is an important result of this paper that screening and form factor effects limit the number of protons that one can coherently scattered from, for any one momentum transfer.

\subsection{Shear viscosity results}
\label{viscosresults}
  
We use results for $S_p(q)$ in Figs. \ref{Fig4},\ref{Fig5},\ref{Fig6} to calculate the Coulomb logarithm $\Lambda_{ep}$, in Eq. \ref{coullog}, and the viscosity $\eta$, from Eq. \ref{eta}.  These results are collected in Table \ref{Table2}.  To gain insight into the results we approximate $S_p(q)$ in Eq. \ref{coullog} with a rectangle of height $Z^*$, width $\Delta q$, and centered at $q^*$ so that $\Lambda_{ep}\approx \Delta q Z^*/q^*$.  This yields a very simple approximation for $\eta$,
\begin{equation}
\eta\approx \frac{\pi n_e}{20\alpha^2} \, \Bigl(\frac{q^*}{\Delta qZ^*}\Bigr)\, .
\label{etaapprox}
\end{equation}
Therefore, $\eta$ follows from the parameters $Z^*$, $q^*$ and $\Delta q$ that characterize $S_p(q)$.

The position of the peak $q^*$ is expected to increase slightly with density as the clusters are packed more closely together.  Perhaps more interesting is the behavior of the width $\Delta q$ with increasing density.  The width describes the range of momentum transfers over which $e-p$ scattering is effective.  This is limited for low $q$ by ion screening and at high $q$ by the nuclear form factor.  Screening can be effective for wavelengths longer than the distance between ions, while the form factor is small for wavelengths much smaller than the size of a nucleus.  Therefore we have the following very approximate way to think about $\Delta q$.
\begin{equation}
\Delta q \approx \frac{1}{\rm Cluster\ size} - \frac{1}{\rm Distance\ between\ clusters}
\end{equation}
In going from a density of 0.01 to 0.025 fm$^{-3}$ the distance between clusters (ions) decreases.  In addition the cluster size increases as the composition changes from nuclei with $A\approx 100$
to $A\approx 200$.  Thus $\Delta q$ ``gets squeezed from both sides'' and it is smaller at a density of 0.025 than it is at 0.01 fm$^{-3}$.

The effective number of protons $Z^*$ is close to the actual number of protons in a nucleus $Z$ at low densities.  However as the density increases, screening and form factor effects greatly reduce $Z^*$ so that $Z^*\ll Z$.  It appears that simple geometric constraints keep $Z^*$ relatively small even if $Z$ is large.  Indeed Table \ref{Table2} shows that $Z^*$ decreases as one goes from $\rho=0.025$ to 0.05 fm$^{-3}$ and the system changes from isolated nuclei to very long pasta shapes that contain large numbers of protons.

Our results for $\eta$ are collected in Table \ref{Table2}.  Note that these values use the full integration in Eq. \ref{coullog} instead of the approximate form in Eq. \ref{etaapprox}.  However the difference is small.  Chugunov et al. \cite{chugunov}, in calculations for spherical nuclei, find the same order of magnitude for $\eta$.  However they only present results for smaller temperatures and proton fractions.  This confirms our finding that non-spherical pasta shapes do not greatly change the shear viscosity. 

There can be large uncertainties in the pasta shapes or even the densities where pasta phases are present.  How might these uncertainties impact $\eta$?  The viscosity, for whatever complex shapes are present, can be calculated in terms of the simple effective parameters $Z^*$, $\Delta q$, and $q^*$. 

We comment on our use of the Born approximation to describe e-ion scattering.  Second or higher order Born corrections will modify our results somewhat, see for example \cite{born2}, but should not change them qualitatively.  Note, there is some ambiguity in determining the best value of the nuclear chage $Z$ to use in calculating Coulomb distortions in the pasta phase.  Perhaps one could use $Z^*$.

We now discuss the temperature dependence of $\eta$.  We have performed calculations at the relatively high temperature of $T=1$ MeV, because this was the temperature of our MD simulations in ref. \cite{pasta2}.  In general we are interested in $\eta$ for lower temperatures.  Equation \ref{etaapprox} shows that $\eta$ depends on $T$ only through the temperature dependence of the parameters $Z*$, $\Delta q$, and $q*$.  We expect the physical size of the pasta clusters to be similar at lower $T$ to their size at $T=1$ MeV.  However, screening will be more efficient at lower $T$.  This could decrease $Z*$ and somewhat increase $\eta$.  The temperature dependence of $\eta$ should be studied in future work with MD simulations at lower temperatures.

Finally, we comment on neutrino contributions to $\eta$.  At high temperatures, for example during a supernova, momentum carried by neutrinos should dominate the shear viscosity because of their very long mean free path.  However at low temperatures, neutrinos will no longer remain in equilibrium with matter and their density will be very low.  As a result, $\eta$ will be dominated by the electron contributions.

\subsection{Thermal Conductivity}
\label{thermlcond}
The thermal conductivity $\kappa$ of nuclear pasta can be calculated in an almost identical way to $\eta$.  We assume the conductivity is dominated by the energy carried by electrons and that this is limited by electron-ion scattering.  Following Nandkumar and Pethick \cite{nandkumar} we write,
\begin{equation}
\kappa = \frac{\pi v_F^2 k_F k_B^2 T}{12 \alpha^2 \Lambda_{ep}^\kappa}\, ,
\end{equation}
where the Coulomb logarithm $\Lambda_{ep}^\kappa$ is slightly different from the corresponding one $\Lambda_{ep}$ for the shear viscosity.
\begin{equation}
\Lambda_{ep}^\kappa = \int_0^{2k_F} \frac{dq}{q\epsilon(q)^2} (1-\frac{v_F^2q^2}{4k_F^2}) S_{p}(q)\, .
\label{coullogkappa}
\end{equation}
Note that Eqs. \ref{coullog} and \ref{coullogkappa}, involving the static structure factor $S_p(q)$, should in principle involve an integral over the full dynamical response function $S_p(q,\omega)$, see for example ref. \cite{nandkumar}.  In addition, Eq. \ref{coullogkappa} omits a term involving the square of the excitation energy $\omega$, \cite{nandkumar}.  This term is small in the classical limit $\omega\ll T$.  We have calculated the dynamical response function for neutrons $S_n(q,\omega)$, using a semiclassical approximation, in ref. \cite{pastaopacity}.  However, we have not yet calculated the proton dynamical response $S_p(q,\omega)$.  This should be calculated in future work and Eq. \ref{coullogkappa} verified.

Table \ref{Table3} presents results for $\Lambda_{ep}^\kappa$ and $\kappa$.  We find that $\kappa$ increases some what with density as $Z^*$ decreases.  To gain qualitative insight, we approximate $S_p(q)$ as a single peak and write,
\begin{equation}
\kappa\approx \frac{\pi k_F k_B^2 T}{12 \alpha^2} \bigl( \frac{q^*}{\Delta q Z^*}\bigr)\, .
\end{equation}
The relatively high thermal conductivity of nuclear pasta may keep the pasta phases at almost the same temperature as the core of the star.  Note that this conductivity is from the electrons.  A strong magnetic field can reduce the electron contribution to the thermal conductivity in directions perpendicular to the field.  In future work we plan to study the nucleon contributions to $\kappa$.  These may be important in the presence of a strong magnetic field.

\begin{table}
\caption{Thermal conductivity $\kappa$ results.  The Coulomb logarithm, Eq. \ref{coullogkappa}, is $\Lambda_{ep}^\kappa$. }
 \begin{ruledtabular}
 \begin{tabular}{ccc}

   $\rho$\ (fm$^{-3}$) &  $\Lambda_{ep}^\kappa$ & $\kappa$\ ( erg/$K$\ cm s) \\

   \hline

0.01 &  16.1 & $2.5 \times 10^{20}$ \\
0.025 &11.7 & $4.7\times 10^{20}$ \\
0.05 & 6.3 & $1.1\times 10^{21}$ \\

\label{Table3}
\end{tabular}
\end{ruledtabular}
\end{table}

\section{Summary and Conclusions}
\label{conclusions}
In this paper we calculate the shear viscosity $\eta$ and thermal conductivity $\kappa$ of nuclear pasta using molecular dynamics (MD) simulations of a semiclassical model.  Our model includes Coulomb repulsion and reproduces nuclear saturation.  Furthermore, it is directly applicable at low densities in a phase with spherical nuclei as well as for complex pasta phases.  Finally, the model makes no assumptions about the pasta shapes.  Instead the MD simulations could produce any shapes.

We assume the shear viscosity is dominated by the momentum carried by electrons and that this is limited by electron-proton scattering.  We calculate the static structure factor $S_p(q)$ of the protons from our MD trajectories for simulations involving 40,000 to 100,000 nucleons.  We find a peak in $S_p(q)$ that can be characterized by a height $Z^*$, a width $\Delta q$ and a position $q^*$.  The height $Z^*$ represents the effective number of protons that an electron can scatter from.  This is reduced from the total number of protons in a nucleus or pasta cluster because of form factor and ion screening effects.

The shear viscosity can be approximated
\begin{equation}
\eta\approx \frac{\pi n_e}{20 \alpha^2} \bigl(\frac{q^*}{\Delta q\, Z^*}\bigr)\, ,
\end{equation}
with $n_e$ the electron density.  We find that $\eta$, for nuclear pasta, is somewhat increased over that for spherical nuclei, because $Z^*$ is much less than the total number of protons in a piece of pasta.  However $\eta$ is not increased by orders of magnitude just because of the non-spherical pasta shapes.  This is in contrast to conventional complex fluids where large non-spherical molecules can dramatically increase $\eta$.  For nuclear pasta, the non-spherical shapes only impact $\eta$ through their influence on $S_p(q)$ and the parameters $Z^*$, $\Delta q$, and $q^*$.

Different effective interactions may lead to different sizes and shapes for the nuclear pasta.  Indeed it may be very difficult to predict these accurately.  However one can calculate transport properties such as $\eta$, the thermal conductivity $\kappa$ and the electrical conductivity in terms of $S_p(q)$ and the parameters $Z^*$, $\Delta q$, and $q^*$.  In future work we hope to develop a simple way to estimate these parameters, given a pasta configuration.  In addition we plan to calculate other mechanical properties of nuclear pasta such as the shear modulus, and extend our MD pasta simulations to additional values of proton fraction, temperature and density.

\begin{acknowledgments}
We thank Andrew Steiner and Ed Brown for useful comments.  This work was supported in part by DOE grant DE-FG02-87ER40365 and by Shared University Research grants from IBM, Inc. to Indiana University.
\end{acknowledgments}


\begin{thebibliography}{99}
\bibitem{pulsars}  J. Clemens and R. Rosen, ApJ. {\bf 609} (2004) 340.

\bibitem{chugunov} A. I. Chugunov and D. G. Yakovlev, Astronomy Reports {\bf 49} (2005) 724.

\bibitem{rmodes}  See for example L. Lindblom, B. J. Owen and S. M. Morsink, Phys. Rev. Lett. {\bf 80} (1998) 4843.  N. Andersson, K. D. Kokkotas, Int. J. Mod. Phys. {\bf D10} (2001) 381.

\bibitem{flowers}  E. Flowers and N. Itoh, ApJ. {\bf 206} (1976) 218; ApJ. {\bf 230} (1979) 847.

\bibitem{nandkumar} R. Nandkumar and C. J. Pethick, MNRAS {\bf 209} (1984) 511.

\bibitem{benhar} O. Benhar and M. Valli, Phys. Rev. Lett. {\bf 99} (2007) 232501.

\bibitem{cooling} R. Wijnands et al., astro-ph/0405089.  E. M. Cackett et al., MNRAS {\bf 372},479.  R. E. Rutledge et al.,  ApJ. {\bf 580} (2002) 413.  P. S. Shternin et al., arxiv:0708.0086.

\bibitem{thermalcon} C. J. Horowitz, O. L. Caballero, and D. K. Berry, Arxiv:0804.4409.

\bibitem{superbursts} A. Cumming and L. Bildsten, ApJ. {\bf 559} (2001) L127.  T. E. Strohmayer and E. F. Brown, ApJ. {\bf 566} (2002) 1045.  A. Cumming et al., ApJ. {\bf 646} (2006) 429.

\bibitem{pasta} D. G. Ravenhall, C. J. Pethick, and J. R. Wilson, Phys. Rev. Lett. {\bf 50} (1983) 2066.  M. Hashimoto, H. Seki, and M. Yamada, Prog. Theor. Phys. {\bf 71} (1984) 320.

\bibitem{pastaopacity} C. J. Horowitz, M. A. Perez-Garcia, D. K. Berry and J. Piekarewicz, Phys. Rev. {\bf C72} (2005) 035801.

\bibitem{pasta1} C. J. Horowitz, M. A. Perez-Garcia, and J. Piekarewicz, Phys. Rev. {\bf C69} (2004) 045804.

\bibitem{pasta2} C. J. Horowitz, M. A. Perez-Garcia, J. Carriere, D. K. Berry, and J. Piekarewicz, Phys. Rev. {\bf C70} (2004) 065806.

\bibitem{verlet} L. Verlet, Phys. Rev. {\bf 159} (1967) 98.

\bibitem{dielectric} B. Jancovici, J. Stat. Phys. {\bf 17} (1977) 357.

\bibitem{hetero} O. L. Caballero, C. J. Horowitz, and D. K. Berry, Phys. Rev. {\bf C74} (2006) 065801.

\bibitem{sawyers} R. F. Sawyer, Phys. Lett. {\bf B630} (2005) 1. 

\bibitem{born2} N. Itoh, S. Uchida, Y. Sakamoto, and Y. Kohyama, arxiv:0708.2967.

\end{thebibliography}

\end{document}